\documentclass[aps,prb,reprint,groupedaddress,showpacs,superscriptaddress,
]{revtex4-1}
\usepackage{hyperref}
\usepackage{amssymb}
\usepackage{graphicx}
\usepackage{dcolumn}
\usepackage{bm}
\usepackage{amsmath}
\usepackage{fixmath}
\usepackage[usenames]{color}
\usepackage{float}
\usepackage{bbm}
\usepackage{epsfig}
\newcommand{\beq}{\begin{equation}}
\newcommand{\eeq}{\end{equation}}
\newcommand{\beqa}{\begin{eqnarray}}
\newcommand{\eeqa}{\end{eqnarray}}
\newcommand{\ba}{\begin{align}}
\newcommand{\ea}{\end{align}}

\newcommand{\dc}{\sigma_{\text{dc}}}
\newcommand{\sgmin}{\sigma_{\text{min}}}
\newcommand{\sv}{\sigma_{0}}
\newcommand{\bk}{\bm{k}}
\preprint{\fbox{\texttt{\tiny\jobname.tex}}}
\begin{document}
\title{DC conductivity of graphene with disorder
}
\author{Michael Sentef}
\email[]{sentefmi@stanford.edu}
\affiliation{Theoretical Physics III, Center for Electronic Correlations and 
Magnetism, Institute of Physics, University of Augsburg, D-86135 Augsburg,
Germany}
\affiliation{Stanford Institute for Materials and Energy Science,
SLAC National Accelerator Laboratory, 2575 Sand Hill Road, Menlo Park, CA 94025, USA}
\author{Marcus Kollar}
\affiliation{Theoretical Physics III, Center for Electronic Correlations and 
Magnetism, Institute of Physics, University of Augsburg, D-86135 Augsburg,
Germany}
\author{Arno P.\ Kampf}
\affiliation{Theoretical Physics III, Center for Electronic Correlations and 
Magnetism, Institute of Physics, University of Augsburg, D-86135 Augsburg,
Germany}
\date{\today}
\begin{abstract}
We model disorder in graphene by random impurities treated in a coherent-potential approximation. Using the analytically solvable Lloyd model for the disorder distribution, we show that the temperature dependence of the minimum conductivity as well as the temperature dependence of the resistivity at high densities and the density dependence of the respective slopes are consistently explained by a temperature dependent disorder strength $\Gamma$ consisting of a constant plus a $T$-linear contribution. This finding suggests that at least two contributions to scattering in graphene are important for its transport properties, and that one of the contributions is due to scattering of electrons from thermally induced excitations.
\end{abstract}
\maketitle
\section{Introduction}
Two of the hallmark features of graphene are its linear Dirac-cone quasiparticle dispersion and the finite minimum conductivity.\cite{Geim07} In non-suspended graphene (NSG) the minimum conductivity $\sgmin$, i.e., the minimal value of the dc conductivity $\dc$ with respect to variations of the electron density, is weakly dependent on temperature $T$ but varies considerably from sample to sample.\cite{Geim07,Tan07} In the attempts to approach the ballistic limit of Dirac fermions without scattering, realizations of suspended graphene (SG) sheets have been prepared, which enable unprecedented electron mobilities \cite{Morozov08,Bolotin08mob} and show a stronger increase of $\sgmin$ upon increasing $T$.\cite{Du08,Bolotin08} For low electron densities when the Fermi energy is slightly off-center of the Dirac cone, $\dc$ decreases with $T$ below a crossover temperature, while the resistivity increases linearly with $T$ at high densities.\cite{Bolotin08}

Free electrons on the half-filled honeycomb lattice constitute a perfect conductor with a non-zero Drude weight at finite temperature and a finite dc conductivity $\sv=\pi e^2/2h$ at zero temperature.\cite{Ziegler07,Stauber08,Lewkowicz09,Neto09review} This is a direct consequence of the Dirac-cone structure of the electronic dispersion. For finite temperatures also the optical conductivity is of order $\sv$ in the visible frequency range. This theoretical result, neglecting disorder effects, is indeed observed in optical absorption experiments on charge-neutral graphene.\cite{Nair08}

Graphene samples are not pristine, however.\cite{Novoselov05,Zhang05} Scanning-tunneling microscopy images of NSG samples show inhomogeneous patterns;\cite{Martin08} their origin was traced to the presence of impurities at the substrate-graphene interface.\cite{Zhang09} Impurities may nucleate electron- or hole-rich puddles,\cite{Guinea08} which obscure the intrinsic Dirac fermion physics of pristine graphene. In suspended graphene (SG) scattering can occur due to microscopic corrugations of the otherwise unstable two-dimensional crystal, so-called ripples, which were observed by transmission electron microscopy \cite{Meyer07} and theoretically analyzed as one possible source for electron scattering \cite{Fasolino07,Katsnelson08} or charge inhomogeneity.\cite{Brey08} Impurity effects are quantitatively less important for the optical conductivity, but play a major role in determining the dc transport properties of graphene; the latter are the topic of this paper.

$\dc$ of graphene is minimal at the charge neutrality point, which corresponds to a honeycomb lattice at half-filling, i.e., with one electron per lattice site. Upon applying a gate voltage of either sign, $\dc$ increases,\cite{Bolotin08} hence the name ``minimum conductivity''. Disorder-induced charge-density modulations imply a spatially varying chemical potential and thereby conceal clean Dirac fermion physics. Also the dc transport measurement process itself may introduce a bias, e.g., due to a charge transfer at metal contacts.\cite{Blake09}

\begin{figure}[htp!]
\begin{center}
\includegraphics[width=\hsize,angle=0]{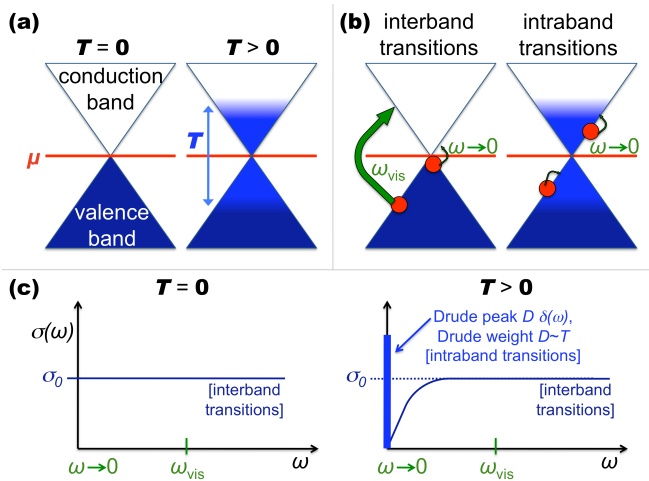} \\
\end{center}
\vspace{-5mm}
\caption[Transport in pristine graphene]{
Transport at the charge neutrality point in pristine graphene. (a) At $T=0$ the valence band (states below the chemical potential $\mu$), shown here for a single Dirac cone, is filled (indicated by the dark blue shading) and the conduction band is empty. At $T>0$ electrons are thermally excited to the conduction band (lighter blue shading). (b) Contributions to the conductivity: Only interband transitions are allowed at $T=0$ due to Fermi blocking. Intraband transitions contribute to dc transport at finite temperatures. (c) Dynamical conductivity of pristine graphene within the Dirac-cone approximation at $T=0$ and $T>0$. At $T=0$ interband transitions lead to a universal finite conductivity of $\sv$ $=$ $\pi e^2/2h$. For $T>0$ a Drude peak emerges due to the intraband transitions with a Drude weight $D$ $\propto$ $T$. For visual frequencies $\omega_{\text{vis}}$ the optical conductivity still is of order $\sv$.}  
\label{fig:0}
\vspace{-3mm}
\end{figure}
Theoretical work on transport in graphene comprises studies of charged-impurity scattering, as reviewed in Ref.\ \onlinecite{Adam09review}, different sources of disorder,\cite{Adam10review} and also the crossover between low- and high-density regimes.\cite{Adam09rapcomm} More recent studies focused on ballistic \cite{Mueller09} or diffusive transport,\cite{Adam09preprint} and also the effects of finite-range scattering at finite densities.\cite{Ferreira11}
Unresolved problems remain in particular in the low-density regime, which is relevant for the minimum conductivity at zero bias. 

 Several predictions exist for a minimum conductivity of
 $4e^2/\pi h$ in the absence of disorder.\cite{theories1,theories2,theories3,theories4,theories5} This limiting value at zero temperature is obtained if the dc limit is taken first and the zero-disorder limit afterwards.\cite{Ziegler07,Lewkowicz09} Experiments on both NSG and SG samples \cite{Geim07,Bolotin08,Du08} with non-universal values of the minimum conductivity were reported, with the trend that the $T$ dependence of the minimum conductivity is enhanced in clean SG samples as opposed to dirty SG \cite{Bolotin08} or NSG samples.\cite{Bolotin08,Du08} In fact, the minimum conductivity increases with increasing temperature, i.e., as in a semiconductor. In contrast, at sufficiently large gate voltages a metallic $T$ dependence of the conductivity is observed, with a resistivity increasing linearly with $T$ and a slope that decreases upon an increase in the gate voltage.\cite{Bolotin08}

Here we evaluate the Kubo formula for the dc conductivity of electrons with a linear Dirac cone dispersion. Disorder effects are included by a random chemical potential, which is treated within the coherent-potential approximation (CPA).\cite{Taylor67,Soven67} The associated disorder energy scale $\Gamma$ may itself depend on temperature. We specifically investigate the case of a Lorentzian disorder distribution of width $\Gamma$ (``Lloyd model''), for which the Kubo formula can be evaluated exactly within CPA. As a result $\dc$ at half-filling depends only on the type of disorder distribution and the dimensionless ratio $T/\Gamma$. For $T/\Gamma$ $\gg$ 1 the minimum conductivity increases linearly with $T/\Gamma$. For a temperature dependent $\Gamma$ $=$ $\Gamma_0 + \alpha_1 T$ the  minimum conductivity thus saturates at high temperatures. With this simple ansatz and a choice of typical meV energy scales for $\Gamma$, the $T$ dependence of $\dc$ changes from semiconducting at half-filling to metallic at sufficiently large band filling. At intermediate densities $\dc$ evolves from metallic to semiconducting behavior in the temperature range between 0 and 200 K. Moreover, the experimentally observed linearly increasing resistivity at high temperatures in the metallic regime as well as the decreasing slope upon increasing the density are reproduced in this ansatz.

\section{Model and Method}
An infinite sheet of pristine graphene is modeled by a tight-binding Hamiltonian with an effective next nearest neighbor hopping on a honeycomb lattice without impurity scattering and electron-electron interactions. In the absence of current-vertex corrections the dc conductivity follows from 
\beq
\dc = \frac{2\pi e^2}{\hbar^2} \int\limits_{-\infty}^{\infty} \hspace{-1mm} \text{d}\nu  \hspace{-1mm} \int\limits_{-\infty}^{\infty}  \hspace{-1mm} \text{d}\epsilon \; \tilde{\rho}(\epsilon) 
\left[ A_{\epsilon}(\nu)+A_{-\epsilon}(\nu) \right]
A_{\epsilon}(\nu)\frac{-\text{d} f_{\nu-\mu}}{\text{d} \nu},
\label{Kubo}
\eeq
where $f_{x}$ $=$ $1/(1+\exp(x/T))$ is the Fermi-Dirac distribution function (with $k_B$ $=$ $1$) and $\tilde{\rho}(\epsilon)$ $=$ $L^{-1}\sum_{\bk}  (\partial\epsilon_{\bk}/\partial k_x)^2 \delta(\epsilon-\epsilon_{\bk})$; $L$ is the number of unit cells of the lattice. $\mu$ is the chemical potential which vanishes at half-filling. In Eq.~(\ref{Kubo}) a prefactor of 4 has been incorporated; it accounts for the spin and valley degeneracies of graphene.

For free electrons the spectral functions simply reduce to $A_{\epsilon}(\nu)$ $=$ $\delta(\epsilon-\nu)$. We use the Dirac cone approximation $\tilde{\rho}(\epsilon)$ $=$ $\hbar|\epsilon|/2\pi$ for $|\epsilon|$ $<$ $\epsilon_{\text{max}}$, where $\epsilon_{\text{max}}$ is a cutoff energy chosen as the half-bandwidth of graphene. Indeed, the Dirac cone approximation for $\tilde{\rho}(\epsilon)$ gives the correct result for $\dc$ and serves as a good approximation even in the visual frequency range,\cite{Stauber08} where the band dispersion leads to only weak quadratic corrections to $\sigma(\omega)$ at low frequencies.

The term in Eq.\ (\ref{Kubo}) which involves $A_{\epsilon}(\nu)^2$ in the integrand leads to the usual intraband conductivity as in single-band models. It gives rise to a Drude-like contribution, hence an infinite dc conductivity in a perfect conductor. Also in the presence of electron-electron interactions this expression for the intraband conductivity remains correct, if the self-energy $\Sigma(\nu)$ is local and $A_{\epsilon}(\nu)$ $=$ $-\text{Im}(\epsilon-\nu-\Sigma(\nu))^{-1}/\pi$.\cite{opticaldmft1,opticaldmft2} The second term in Eq.\ (\ref{Kubo}), involving $A_{\epsilon}(\nu) A_{-\epsilon}(\nu)$, describes excitations with a particle at energy $\epsilon$ and a hole at $-\epsilon$ and accounts for interband transitions. This contribution accounts for the visual transparency of graphene in the dc limit of the optical conductivity, $\sv$ $=$ $\pi e^2/2h$ (see Fig.\ \ref{fig:1}c).\cite{Nair08}

The presence of disorder complicates the situation considerably. Discrete translational invariance is broken, rendering microscopic theoretical approaches much more difficult than in the homogeneous case. One standard approach is the Anderson model \cite{Anderson58} with local potential impurities,
\beq
H = H_0 + \sum_i V_i n_i,
\eeq
where $H_0$ is the tight-binding Hamiltonian for the clean system, $n_i$ is the local density operator on site $i$ of the lattice and $V_i$ is a random variable determined from a probability distribution $P(V_i)$. Here we do not aim at a full microscopic description of disorder, e.g., in the spirit of a self-consistent diagrammatic treatment of impurity scattering effects,\cite{Vollhardt80} and recall that weak localization is suppressed by long-range scattering in graphene.\cite{Morozov06, Wakabayashi07, Ziegler08} Instead we apply the coherent-potential approximation (CPA)\cite{Taylor67,Soven67,Elliott74} to determine an effective random medium described by a local self-energy $\Sigma(\omega)$, which is determined by a self-consistent solution of the CPA equations
\beqa
\bar{G}(\omega) &=& G_0(\omega-\Sigma(\omega)),\nonumber\\
\bar{G}(\omega) &=& \tilde{D}\left[\mathcal{G}^{-1}(\omega)\right] = \int \text{d}V\; \frac{P(V)}{\mathcal{G}^{-1}(\omega)-V},\nonumber\\
\mathcal{G}^{-1}(\omega) &=& \bar{G}^{-1}(\omega) + \Sigma(\omega).
\label{CPA}
\eeqa
In Eq.~(\ref{CPA}) $G_0(z)=\int \text{d}\omega\; \rho_{\text{DOS}}(\omega)/(z-\omega)$ is the local Green function of the clean system described by $H_0$, $\mathcal{G}(\omega)$ is a dynamical Weiss field and $\tilde{D}[z]$ is the Hilbert transform with respect to the disorder distribution function $P(V_i)$.  The CPA expression for the conductivity \cite{Elliott74} agrees with the Kubo formula Eq.\ (\ref{Kubo}) with $A_{\epsilon}(\nu)$ $=$
 $-\text{Im}(\epsilon-\nu-\Sigma(\nu))^{-1}/\pi$.

The CPA equations (\ref{CPA}) can be solved, at least numerically, for an arbitrary disorder distribution. In order to keep the subsequent analysis as simple and transparent as possible, we focus on the specific case of a Lorentzian disorder distribution of width $\Gamma$,
\beq
P(x) = \frac1\pi \frac{\Gamma}{\Gamma^2+x^2}.
\eeq
This is the so-called Lloyd model, for which the CPA equations are exactly solvable using $\tilde{D}[z]$ $=$ $(z+i\Gamma)^{-1}$, which yields $\Sigma(\nu)$ $=$ $-\text{i}\Gamma$. Hence we obtain for the Lloyd model
\beq
A_{\epsilon}(\nu) = \frac1\pi \frac{\Gamma}{\Gamma^2+(\epsilon-\nu)^2}
\eeq
as the input quantity for Eq.\ (\ref{Kubo}). $A_{\epsilon}(\nu)$ is of the form $a((\epsilon-\nu)/\Gamma)/\Gamma$, implying that the dc conductivity is only a function of the ratio $T/\Gamma$ for $\mu=0$ and of $\mu/\Gamma$ for $T=0$; the latter holds only, if $\mu$ $\ll$ $\epsilon_{\text{max}}$, which is fulfilled in the experiments cited above. 

The scaling behavior of $\sgmin$ has two reasons: (a) the cutoff energy $\epsilon_{\text{max}}$ can be replaced by infinity in the $\epsilon$-integral in Eq.\ (\ref{Kubo}) and thus does not appear as an additional energy scale, and (b) for dimensional reasons the dc conductivity is universal in the sense that it does not depend on the hopping matrix element $t$ of the underlying two-dimensional tight-binding Hamiltonian and, consequently, not on the Fermi velocity $v_F$. Both reasons are directly related to the linearity of the dispersion in graphene up to energies much larger than the relevant temperatures. For finite densities this universality no longer holds, since the density variations are determined by the chemical potential which thereby depends on the hopping matrix element $t$.

Experimentally it is the gate voltage which controls the electronic density $n$, measured relative to half-filling. For given temperature $T$, disorder strength $\Gamma$, and chemical potential $\mu$ the density is given by
\beq
n = \int_{-\infty}^{\infty} \text{d}\omega\; \rho_{\text{DOS}}(\omega) \left(f_{\omega-\mu}-f_{\omega}\right),
\eeq
where
\beq
\rho_{\text{DOS}}(\omega) = \frac{4}{\sqrt{3}\pi t^2 A_u} \int_{-\epsilon_{\text{max}}}^{\epsilon_{\text{max}}}\text{d}\epsilon\; |\epsilon| A_{\epsilon}(\omega)
\eeq
is the density of states (summed over both spin projections) for the disordered system, $t$ $=$ 2.7 eV the hopping matrix element of the tight-binding model, $A_u$ $=$ $3\sqrt{3}a_0^2/2$ the size of the unit cell, and $a_0$ $=$ 1.42 $\times$ $10^{-10}$ m the interatomic distance on the honeycomb lattice.\cite{Neto09review}

\section{Results}
We first keep the disorder strength $\Gamma$ fixed and discuss basic properties of the minimum conductivity and the conductivity at finite chemical potential and zero temperature. In a second step we evaluate the density dependence of the conductivity for a typical disorder strength (on the order of meV\cite{Du08,Martin08}) and for temperatures in the range from 0 K to 200 K. Especially we consider the temperature dependence of the resistivity $\dc^{-1}$ at fixed densities and show that a temperature-independent $\Gamma$ at high densities is insufficient to explain the experimentally observed linear $T$ dependence of the resistivity. Adding a phenomenological linear $T$ dependent contribution to $\Gamma$, the experimental observation is matched by our ansatz. Moreover, the observed density dependence of the slope in the $T$-linear regime of the resistivity follows naturally without further assumptions. For a selected $T$ dependent $\Gamma$ $=$ $\Gamma_0 + \alpha_1 T$, the minimum conductivity increases with temperature but with a decreasing slope; $\sgmin$ saturates at high temperatures closely similar to the experiments. Finally we show the $T$ dependence of the conductivity with a density dependent crossover from metallic ($\text{d}\dc/\text{d}T$ $<$ 0) at low $T$ to semiconducting behavior ($\text{d}\dc/\text{d}T$ $>$ 0) at high $T$.

\begin{figure}[htp!]
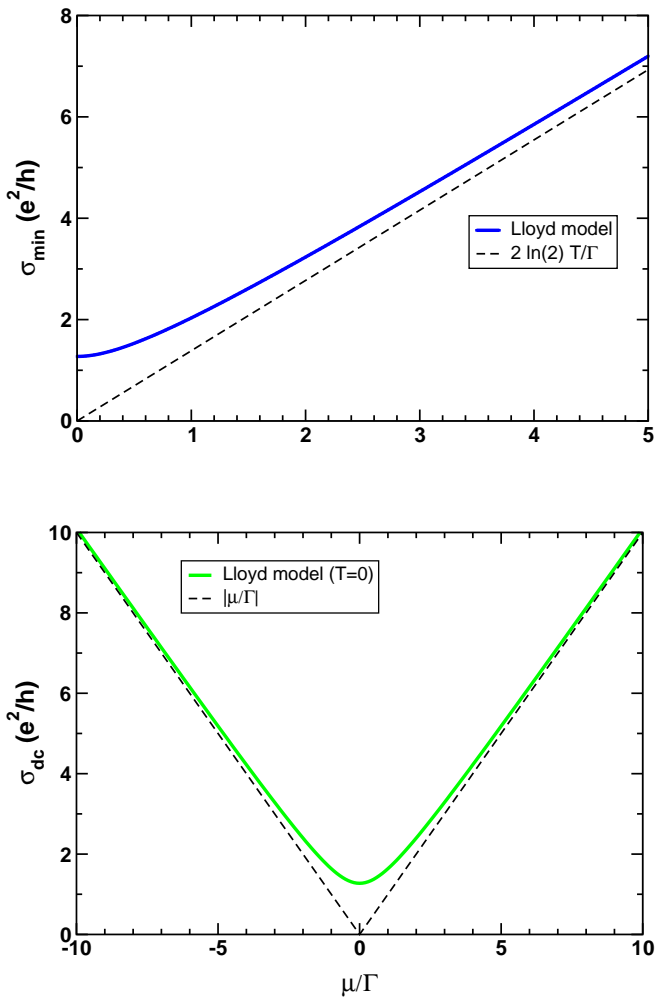

\includegraphics[width=\hsize,angle=0]{sgmin_lloyd.eps} \\
\vspace{5mm}
\includegraphics[width=\hsize,angle=0]{sgnu_lloyd.eps}
\caption[DC conductivity for the Lloyd model]{
DC conductivity in units of $e^2/h$ for the Lloyd model in CPA. Top panel: Minimum conductivity ($\mu$ $=$ 0) as a function of $T/\Gamma$. Bottom panel: DC conductivity at $T$ $=$ 0 as a function of $\mu/\Gamma$.}
\label{fig:1}
\end{figure}
The temperature dependence of the minimum conductivity $\sgmin$ for the Lloyd model with disorder strength $\Gamma$ is shown in the top panel of Fig.\ \ref{fig:1}. For $T/\Gamma$ $\rightarrow$ 0, $\sgmin$ tends to the limiting value $4e^2/\pi h$, which coincides with the clean limit discussed in Refs.\ \onlinecite{theories1,theories2,theories3,theories4,theories5}. However, $4e^2/\pi h$ should not be considered a universal value, but rather a particular result of the Lloyd model. Other non-Lorentzian disorder distributions are likely to lead to other values of the minimum conductivity. At high temperatures $\sgmin$  increases linearly with $T/\Gamma$ for $T/\Gamma$ $\gg$ 1, 
\beq
    \sgmin = \frac{e^2}{h}
    \bigg(2 \ln(2)\,\frac{T}{\Gamma}
            +{\cal O}\bigg(\frac{\Gamma}{T}\bigg)
    \bigg)
    \,.\label{hightemp}
\eeq

To understand the physical processes involved we discuss the relevant contributions to $\sgmin$. First we note that the clean case at zero temperature is not recovered by our theory for $\dc$. However, this is not a shortcoming but rather a generic feature of the conductivity as a function of disorder strength, temperature, and frequency. We recall that at zero temperature intraband excitations are prohibited. Thus only the interband excitations are responsible for $\text{Re}\;\sigma(\omega \rightarrow 0)$ $=$ $\sv$ \cite{Stauber08} (see Fig.\ \ref{fig:0}), i.e. when the dc limit is taken \emph{after} the limits of zero temperature and zero disorder strength. The theory presented here instead aims at describing dc measurements, for which the dc limit must be taken \emph{first}. In the latter case, both interband and intraband excitations are relevant and both contribute equally ($2e^2/\pi h$ for the Lloyd model) to the $T/\Gamma$ $\rightarrow$ 0 limit. The discrepancy between $\sv$ and $\sgmin(T \rightarrow 0)$ may also be understood by noting that for $\sv$ the largest energy scale in the system is the frequency (taken to zero last), while the largest energy scale for $\sgmin$ is the disorder strength $\Gamma$.

At finite temperatures thermally excited particles in the conduction band render intraband particle-hole excitations possible, leading to a non-zero Drude weight and thus an infinite $\dc$ (for $\Gamma$ $\rightarrow$ 0 and therefore $T/\Gamma$ $\rightarrow$ $\infty$), while interband low-energy excitations are blocked by thermally occupied states in the conduction band. Viewed as a function of temperature at fixed $\Gamma$ the conductivity is semiconducting, i.e., $\text{d}\dc/\text{d}T$ $>$ 0.

At finite chemical potentials the interband excitations become less important, and the behavior is determined mostly by intraband excitations. The situation for higher densities thus resembles more and more the case of a single partially filled band, where metallic behavior sets in for sufficiently low temperatures, i.e., $\text{d}\dc/\text{d}T$ $<$ 0.

The dc conductivity at $T$ $=$ 0 as a function of $\mu/\Gamma$ is shown in the lower panel of Fig.\ \ref{fig:1}. It tends to the limiting value $4e^2/\pi h$ for $|\mu/\Gamma|$ $\rightarrow$ 0 and increases linearly for $|\mu/\Gamma|$ $\gg$ 1 but well below the cutoff $\epsilon_{\text{max}}/\Gamma$. This linear dependence on the chemical potential for large $|\mu/\Gamma|$ is analytically obtained from the Kubo formula, taking into account intraband excitations only,
\beqa
\dc(\mu,T=0) &\approx& \frac{2e^2}{\pi h} \int_{-\infty}^{\infty} \text{d}\epsilon |\epsilon| \frac{\Gamma^2}{\left(\Gamma^2+(\epsilon-\mu)^2\right)^2}
\nonumber
\\
&=&
\frac{2e^2}{\pi h} \left( 1 + \frac{\mu}{\Gamma} \arctan\frac{\mu}{\Gamma} \right)
\sim
\left|\frac{\mu}{\Gamma}\right| \frac{e^2}{h},
\eeqa
where the last asymptotic expression is valid for $|\mu/\Gamma|$ $\gg$ 1.

\begin{figure}[htp!]
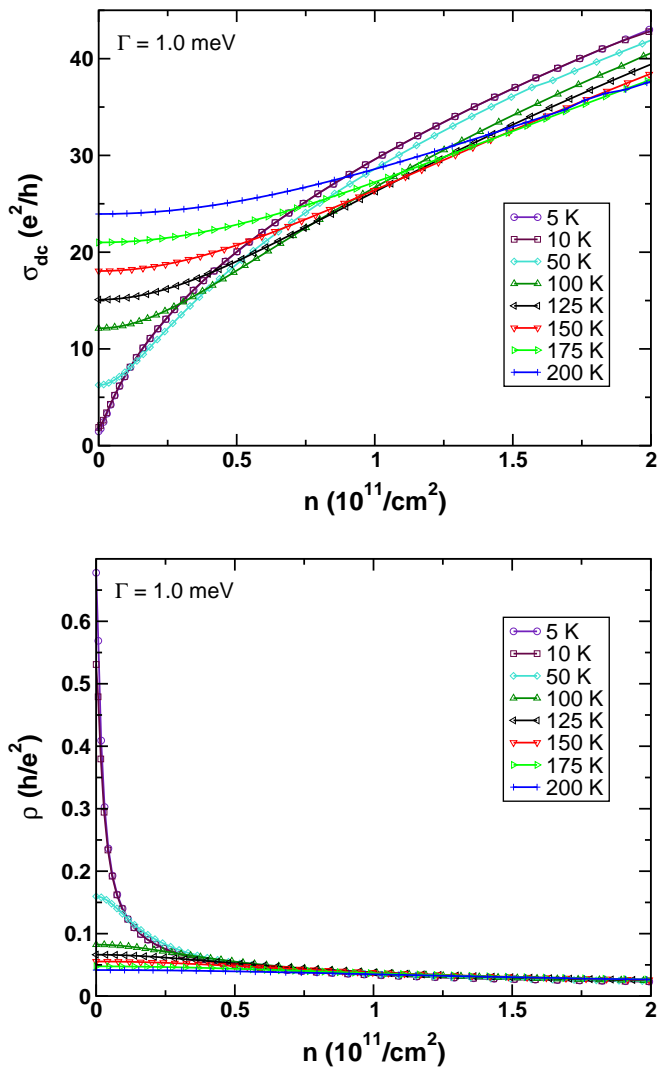

\begin{center}
\includegraphics[width=\hsize,angle=0]{sgn_lloyd.eps} \\
\vspace{5mm}
\includegraphics[width=\hsize,angle=0]{rhon_lloyd.eps}
\end{center}
\vspace{-5mm}
\caption[Density-dependent conductivity and resistivity]{
Upper panel: Conductivity as a function of density at different temperatures for a temperature independent $\Gamma$. Lower panel: The data for the resistivity $\rho$ $=$ $\dc^{-1}$.}
\label{fig:2}
\vspace{-3mm}
\end{figure}
For fixed value of the disorder strength $\Gamma$ $=$ 1 meV we show in Fig.\ \ref{fig:2} the dc conductivity and its inverse, the resistivity $\rho$, as a function of density for selected temperatures. Since the density depends quadratically on the chemical potential for $\mu/\Gamma$ $\gg$ 1 and $\dc$ depends linearly on $|\mu/\Gamma|$ in this limit at zero temperature, the low-temperature conductivity increases like $\sqrt{n}$ at high densities. Experimentally a sublinear density dependence of the conductivity was also reported in Ref.\ \onlinecite{Du08}. 

In Ref.\ \onlinecite{Bolotin08} a linear increase of the resistivity as a function of temperature was observed for high densities at elevated temperatures. For a temperature independent $\Gamma$, $\dc$ $=$ $|\mu/\Gamma|$ $e^2/h$ for large $|\mu/\Gamma|$, and the temperature dependence of the chemical potential at fixed densities follows $\mu$ $=$ $\mu(T=0)+\mathcal{O}(T^2)$. Hence, $\dc(T)$ $=$ $\dc(T=0)+\mathcal{O}(T^2)$ for fixed densities and thus $\rho$ $=$ $\rho(T=0)+\mathcal{O}(T^2)$. A linear temperature dependence of $\rho$ therefore requires a temperature dependent $\Gamma$ within our ansatz. In fact, a linearly increasing resistivity at high densities naturally follows from $\Gamma$ $=$ $\Gamma_0+\alpha_1 T$,
\beq
\rho \approx \frac\Gamma\mu \frac{h}{e^2} = \frac{\Gamma_0}{\mu}\frac{h}{e^2} + \frac{\alpha_1 T}{\mu}\frac{h}{e^2}.
\label{Tdep}
\eeq
$\mu$ thereby depends not only explicitly on temperature, but also implicitly via the $T$ dependent $\Gamma$. This implicit $T$ dependence is, however, negligible for large fillings, when also the explicit $T$ dependence is very weak since it scales like temperature over Fermi energy. 

\begin{figure}[htp!]
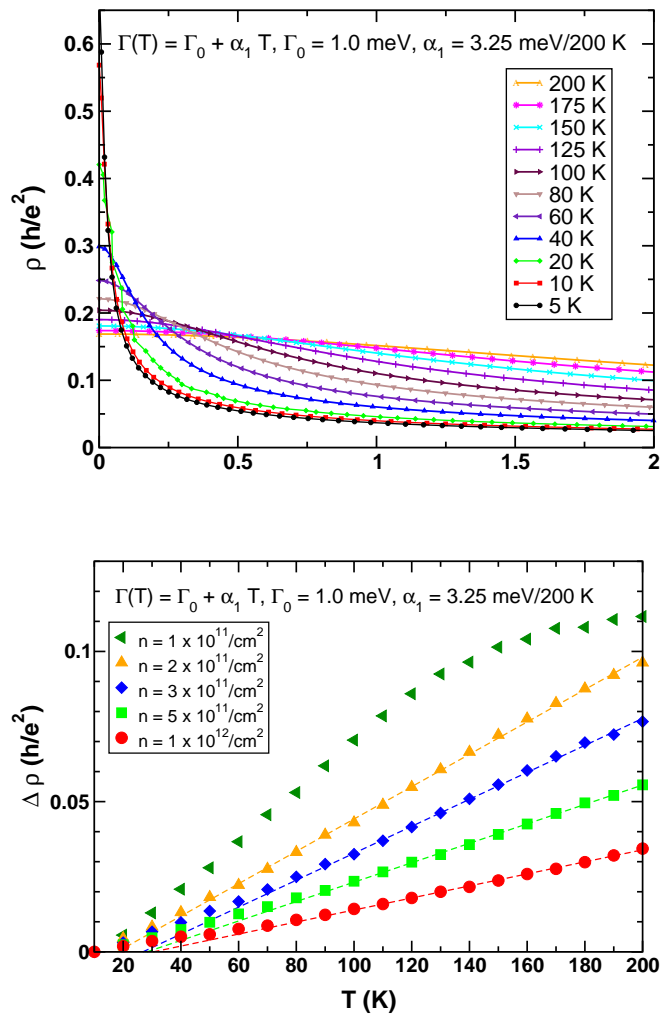

\begin{center}
\includegraphics[width=\hsize,angle=0]{rhon.eps} \\
\vspace{5mm}
\includegraphics[width=\hsize,angle=0]{rhot3.eps} 
\end{center}
\vspace{-5mm}
\caption[Density- and temperature-dependent resistivity]{
Upper panel: Resistivity as a function of density for different temperatures.
Lower panel: Resistivity increase $\Delta \rho$ $=$ $\rho(T)-\rho(\text{10 K})$ above 10 K as a function of temperature for different densities. The dashed lines are linear fits to the data points between 100 K and 200 K.}
\label{fig:3}
\vspace{-3mm}
\end{figure}
\begin{figure}[htp!]
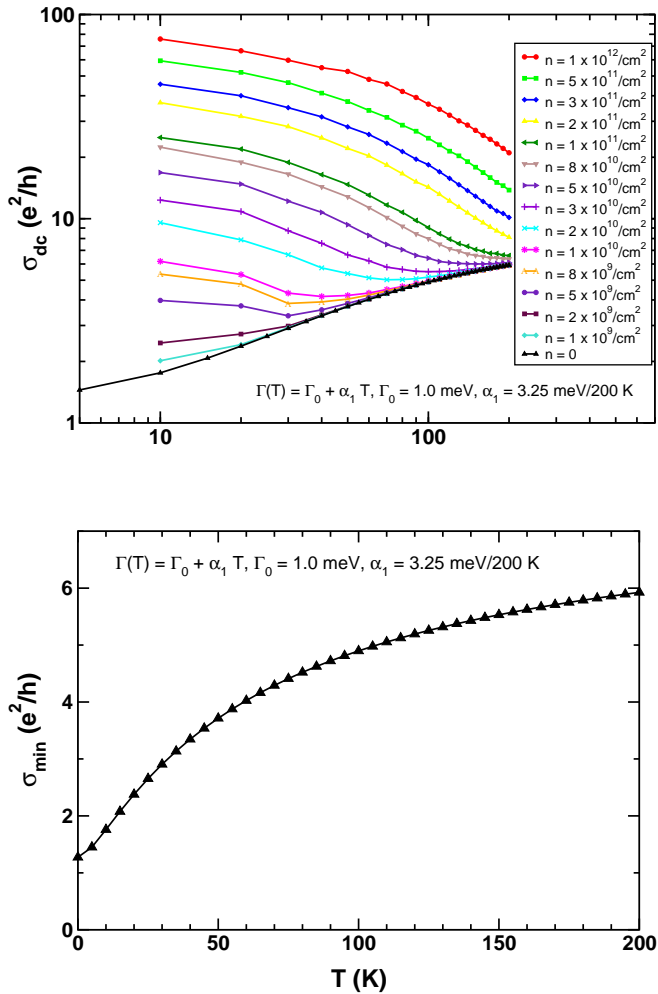

\vspace{6mm}
\begin{center}
\includegraphics[width=\hsize,angle=0]{sgt.eps} \\
\vspace{5mm}
\includegraphics[width=\hsize,angle=0]{sgmin.eps}
\end{center}
\vspace{-5mm}
\caption[Temperature-dependent conductivity]{
Upper panel: Conductivity as a function of temperature for different densities with a $T$ dependent $\Gamma$ in a double-logarithmic scale. The density is measured with respect to half-filling. Lower panel: Temperature dependence of the minimum conductivity.}
\label{fig:4}
\vspace{-3mm}
\end{figure}

In the following we adopt the $T$ dependent disorder strength $\Gamma(T)$ $=$ $\Gamma_0$ $+$ $\alpha_1$ $T$ and fix the parameters $\Gamma_0$ $=$ 1 meV and $\alpha_1$ $=$ 3.25 meV/200 K such that the temperature dependence of the minimum conductivity (see lower panel of Fig.\ \ref{fig:4}) approximately matches the experimental data of Ref.\ \onlinecite{Du08}. The density dependence of the resistivity for the selected $T$ dependent disorder strength is shown in the upper panel of Fig.\ \ref{fig:3}. For moderate temperatures below 200 K the temperature dependence of the resistivity is presented in the lower panel of Fig.\ \ref{fig:3}. Eq.\ (\ref{Tdep}) implies that the slope in the linear regime is proportional to $\alpha_1/\mu$, and since $\mu$ $\propto$ $\sqrt n$ the slope decreases like $1/\sqrt{n}$. For high temperatures well above 200 K and sufficiently large densities, or below 200 K for moderate densities, there is a deviation from linear behavior, and the resistivity decreases again due to interband excitations.

The crossover from metallic behavior at finite densities and low temperatures to semiconducting behavior at elevated temperatures is shown in the upper panel of Fig.\ \ref{fig:4}. The crossover temperature vanishes at zero density ($\mu$ $=$ 0), since $d\sgmin/dT$ $>$ 0 for all temperatures, and also increases with increasing density. In fact, for the selected temperature dependence of $\Gamma$ the conductivity is metallic below 200 K for densities larger than 8 $\times$ 10$^{10}/\text{cm}^2$. The temperature dependence of the minimum conductivity for the same $T$ dependent $\Gamma$ is shown in the lower panel of Fig.\ \ref{fig:4}. Here a sublinear $T$ dependence of $\sgmin$ is observed for elevated temperatures. Indeed, the curvature of $\dc$ changes sign at an intermediate temperature depending on the relative sizes of $\Gamma_0$ and $\alpha_1$. An increasing $\sgmin$ as a function of temperature with a sublinear behavior at elevated $T$, yet below 200 K, is similarly observed in experiments.\cite{Bolotin08,Du08}

\section{Summary and Discussion}
We have presented a phenomenological theory for the temperature dependence of the dc conductivity of graphene at zero and finite particle densities including potential disorder in a coherent-potential approximation (CPA). Specifically we have chosen a Lorentzian disorder distribution (``Lloyd model''), for which the CPA equations are exactly solvable. This approach recovers well-established limits in the clean case and at the same time provides a phenomenological context for the remarkable transport properties of graphene in the presence of impurity scattering. For the Lloyd model the minimum conductivity is $4e^2/\pi h$, which coincides with previous predictions for the dc limit in the clean system provided that the zero frequency limit is taken before the clean limit at zero temperature. At finite temperatures, the enhanced $T$ dependence of the minimum conductivity in cleaner SG samples is explained, and we find $\sgmin$ $\propto$ $T/\Gamma$ for $T/\Gamma$ $\gg$ 1. As a consequence we expect a very steep increase of $\sgmin$ with temperature in even cleaner samples. Moreover we have shown that the $T$ linear resistivity at high densities and the density dependence of its slope follow naturally from a temperature dependent $\Gamma$ $=$ $\Gamma_0 + \alpha_1 T$. This phenomenologically determined $T$ dependence of $\Gamma$ suggests the existence of at least two sources for scattering in suspended graphene. The constant $\Gamma_0$ points to static potential disorder, whereas the $T$-linear part $\alpha_1 T$ may arise from scattering off a thermally excited perturbation. One obvious possibility are thermally excited ripples, since even the linear $T$ dependence of the scattering rate could be explained within the ripple scenario \cite{Katsnelson08}. 

Here we have investigated the role of disorder, as described by an Anderson impurity model with a phenomenological disorder strength, as a source for scattering in graphene. As pointed out in Ref.\ \onlinecite{Blake09}, it is important to understand which additional extrinsic effects may mask the intrinsic properties of graphene, especially the sensitive Dirac fermion physics at the neutrality point. Possible extrinsic perturbations are contact resistances, spurious chemical doping into the contact regions, or macroscopic charge inhomogeneity on length scales comparable to the sample size. Such effects need to be incorporated in order to understand the unusual transport properties of graphene in particular at the charge-neutrality point. Also improved doping techniques using organic molecules \cite{Coletti10} may help to unveil the intrinsic transport properties of grapheme. Further theoretical and experimental activity should clarify these aspects and the promising prospects of graphene as a basis of future electronic devices.

We acknowledge discussions with Prabuddha Chakraborty, Krzysztof Byczuk, Holger Fehske, Andreas Sinner, and Wolfgang H\"ausler. This work was supported by the Deutsche Forschungsgemeinschaft through TRR 80.
\bibliography{mybib}{}
\bibliographystyle{apsrev}
\end{document}